\documentclass[a4,12pt]{article} 
\usepackage{cite}
\usepackage{wrapfig}
\usepackage{graphicx}
\usepackage{amssymb}
\usepackage{amsfonts}
\usepackage{amsmath}
\usepackage{longtable}
\usepackage{rotating}
\usepackage{lscape}
\usepackage{epsfig}
\usepackage{multirow}

\begin{document}

\title{Production of various elements in ultraperipheral $^{208}$Pb--$^{208}$Pb collisions at the LHC}
\author{U.A.\,Dmitrieva$^{a,b,}$\footnote{E-mail: uliana.dmitrieva@phystech.edu}, 
I.A.\,Pshenichnov$^{a,b,}$\footnote{E-mail: pshenich@inr.ru} \\
$^{a}$\,Institute for Nuclear Research \\ 
of the Russian Academy of Sciences, \\
7a~Prospekt 60-letiya Oktyabrya, 117312, Moscow, Russia \\
$^{b}$\,Moscow Institute of Physics and Technology \\ 
(National Research University), \\
9~Institutskiy~per., 141701, Dolgoprudny, Russia
}
\date{}
\maketitle

\begin{abstract}
As predicted by theory and confirmed by measurements, one, two or three neutrons are emitted frequently in ultraperipheral collisions (UPCs) of heavy relativistic nuclei, in particular, $^{208}$Pb. The exchange of low-energy Weizs\"{a}cker--Williams photons dominates in such interactions. This leads to the excitation and decay of Giant Dipole Resonances (GDR) in colliding nuclei below the proton emission threshold. Less is known about the electromagnetic dissociation of $^{208}$Pb induced by energetic photons leading to violent fragmentation of $^{208}$Pb.  The UPCs of lead nuclei at the LHC were modelled with Relativistic ELectromagnetic Dissociation (RELDIS) model to evaluate the contribution of photonuclear reactions in the domain of quasideuteron absorption and at higher photon energies.  It was demonstrated that due to the presence of a single heavy residue in the final state mostly accompanied by free protons and neutrons, the cross sections of the production of specific elements can be well approximated by the proton emission cross sections, which can be measured in the ALICE experiment at the LHC.  
\end{abstract}
\vspace*{6pt}

\label{sec:intro}
\section*{Introduction}

The total cross sections of electromagnetic dissociation (EMD) of lead nuclei calculated for ultraperipheral collisions (UPCs) at the LHC and the future FCC-hh are close to 210~b and 280~b, respectively~\cite{Pshenichnov2019}.  The partial cross sections of emission of 1, 2, 3, 4 and 5 forward neutrons with and without forward protons in UPCs of $^{208}$Pb were measured recently in the ALICE experiment at the LHC at $\sqrt{s_{\rm NN}} = $~5.02~TeV~\cite{Dmitrieva_ALICE_2022}. The data on neutron emission were described by the Relativistic ELectromagnetic DISsociation (RELDIS) model~\cite{Pshenichnov2011} based on the  Weizs\"{a}cker--Williams method~\cite{Bertulani1988} taking into account single and double photon absorption by $^{208}$Pb. Following this method, the Lorentz-contracted Coulomb fields of the colliding nuclei are represented by fluxes of equivalent photons. The impact of the Coulomb field of the nucleus $A_1$ on the nucleus $A_2$ in their UPCs can be considered as the absorption of equivalent photons by $A_2$. Because of the factor $1/E_{\gamma}$ in the spectrum of Weizs\"{a}cker--Williams photons~\cite{Bertulani1988}, the absorption of soft low-energy photons is enhanced leading to the dominance of the excitation and decay of Giant Dipole Resonances (GDR) in colliding nuclei below the proton emission threshold. As shown by calculation with RELDIS~\cite{Dmitrieva_ALICE_2022},  few nucleons and a single residual nucleus are produced in most of EMD events, and the contribution of photofission of $^{208}$Pb is negligible. In the present work the EMD of $^{208}$Pb at the LHC is modeled with the focus on the absorption of high-energy photons leading to the emission of forward protons along with neutrons. The cross sections of the production of various elements other than lead are calculated. The distributions of the total energy of forward neutrons or protons are also considered.

\label{sec:photons_impact}
\section{Production of protons and secondary nuclei by low- and high-energy photons}

As known~\cite{Bertulani1988,Pshenichnov2011}, in the absorption of a low-energy photon ($7<E_\gamma <40$~MeV) the entire target nucleus is excited to a Giant Dipole Resonance (GDR) state. Since the wavelength of low-energy photons is comparable to the size of the nucleus, all protons oscillate relative to neutrons leading to the collective GDR excitation. Photons of higher energies ($40 <E_\gamma < 140$~MeV) with the wavelength comparable to the internucleon distance are absorbed by quasi-deuteron proton-neutron pairs inside the target nucleus. At $E_\gamma > 140$~MeV the photon energy exceeds the pion production threshold and hadron photoproduction takes place on individual nucleons.

\begin{figure}[!htb]
\begin{centering}
\includegraphics[width=1.0\columnwidth]{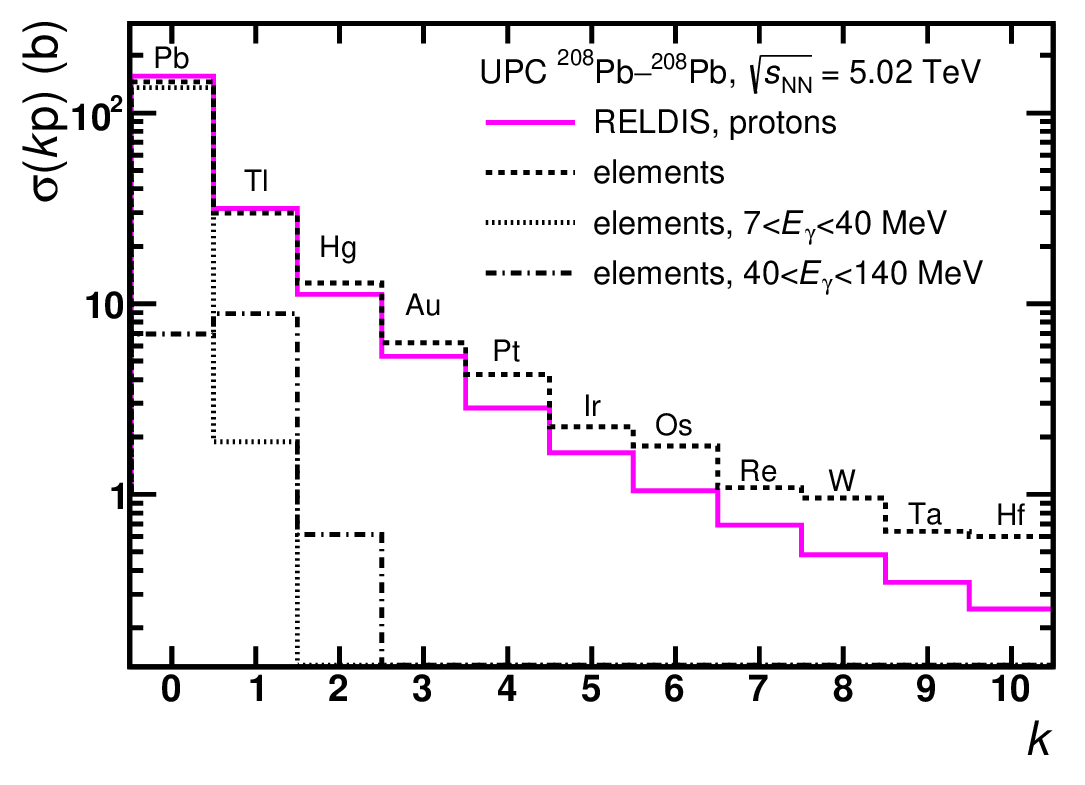}
\caption{The cross sections of emission of a given number of protons (solid histogram) and the cross sections of the production of the corresponding secondary nuclei (dashed histogram) calculated with RELDIS  for UPCs  of  $^{208}$Pb at the LHC at $\sqrt{s_{\rm NN}} = $~5.02~TeV. The contributions of the GDR and quasideuteron absorption are shown by dotted and dashed-dotted histograms, respectively. 
}
\label{fig:sec_nucl_p_Xn}
\end{centering}
\end{figure}

In the present study the EMD of $^{208}$Pb at the LHC were modeled with RELDIS. The cross sections of emission of 1, 2, ... and 10 protons were calculated along with the cross sections of production of Pb, Tl, Hg, Au, Pt, Ir, Os, Re, W, Ta and Hf nuclei, see Fig.~\ref{fig:sec_nucl_p_Xn}. In addition, in dedicated RELDIS runs the contributions to the element production cross sections were calculated specifically from photons of 7--40~MeV and 40--140~MeV representing the GDR and quasideuteron absorption, respectively. As seen in Fig.~\ref{fig:sec_nucl_p_Xn}, the cross sections of production of Pb, Tl, Hg and Au isotopes are well approximated by the corresponding cross sections of the emission of 0, 1, 2 and 3 protons. This is due to relatively low average excitation energy per nucleon delivered to the residual nucleus after the completion of the intranuclear cascade induced by equivalent photons. As explained in Ref.~\cite{Dmitrieva_ALICE_2022}, the nuclear residue evaporates nucleons, but not undergo fission or multifragmentation.  This suggests that by measuring the cross sections of the emission of given numbers of protons, the cross sections to produce the respective elements can be evaluated. As seen in Fig.~\ref{fig:sec_nucl_p_Xn}, almost all Pb nuclei are produced by low-energy photons in the GDR domain, while less than one of 15 Tl nuclei is produced by such soft photons, and other 4--5 nuclei are produced by 40--140~MeV photons. The dominant contribution of energetic photons ($E_\gamma > 140$~MeV) to the spallation reactions  producing other elements (Hg, Au, Pt, Ir, Os, Re, W, Ta and Hf) is also evident.    

\begin{figure}[!htb]
\begin{centering}
\includegraphics[width=1.0\columnwidth]{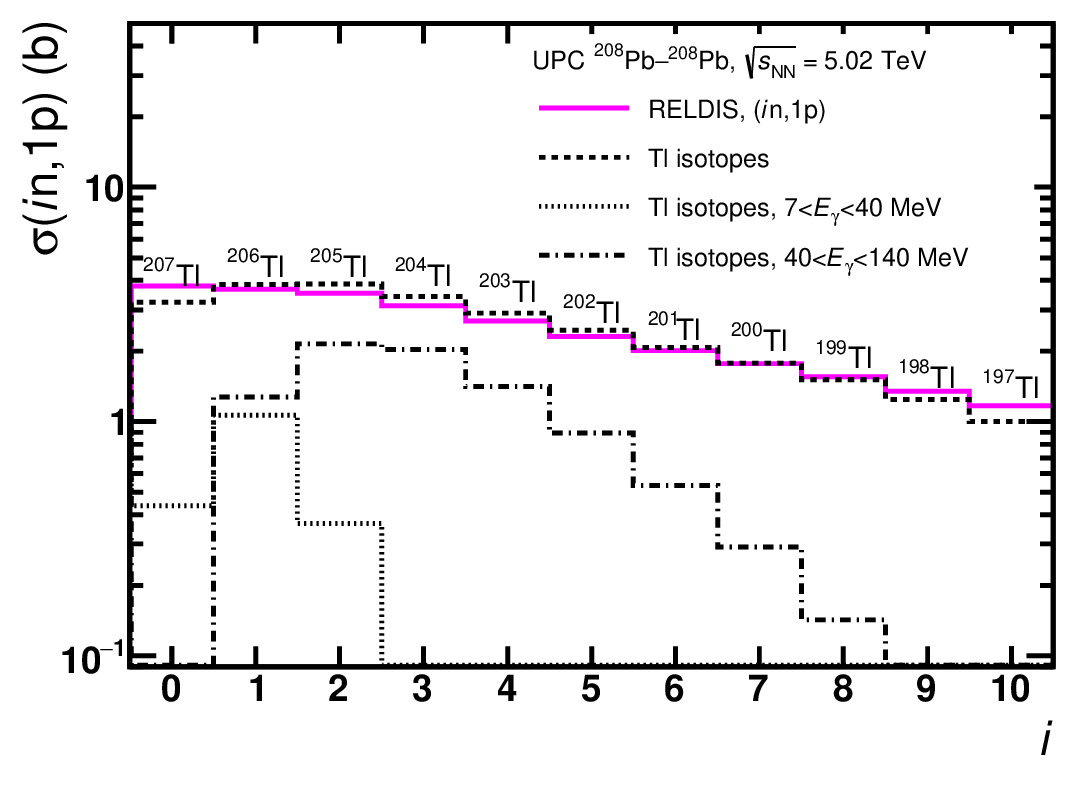}
\caption{The cross sections of emission of a given number of neutrons accompanied by a single proton (solid histogram) and the cross sections of the production of the corresponding Tl isotopes (dashed histogram) calculated with RELDIS  for UPCs  of  $^{208}$Pb  at $\sqrt{s_{\rm NN}} = $~5.02~TeV. The contributions of the GDR and quasideuteron absorption are shown by dotted and dashed-dotted histograms, respectively. 
}
\label{fig:Tl_p}
\end{centering}
\end{figure}

In the reactions induced by energetic photons on $^{208}$Pb, several neutrons and protons are emitted, and isotopes of each element with various mass numbers are produced. This is demonstrated by the calculated cross sections of the production of various thallium isotopes, $^{207, 206, ..., 197}$Tl, presented in Fig.~\ref{fig:Tl_p}. These cross sections are between 1 and 4 barns and comparable to each other. The cross sections of the emission of 0,1,2,... 10 neutrons accompanied by a single proton are also shown in this figure, and they approximate well the cross sections of the production of $^{207,206,...197}$Tl.  As seen in Fig.~\ref{fig:Tl_p}, $\sim 50$\% of $^{204}$Tl and $^{205}$Tl nuclei are produced by photons of $40 <E_\gamma < 140$~MeV, while the contribution of more energetic photons with $E_\gamma > 140$~MeV dominates for other thallium isotopes. It is interesting to note that the heaviest of the considered isotopes, $^{207}$Tl, is produced mainly by knocking out a single proton by energetic photons above 140~MeV  ($\sim 85$\% of events). The rest of events to produce $^{207}$Tl is attributed to the decay of GDR, but the contribution from the quasideuteron absorption is negligible. 

\label{sec:detection}
\section{Distributions of the total energy of EMD neutrons and protons}

In the ALICE experiment at the LHC the detection of secondary nuclei produced in EMD is impossible.  Nevertheless, the total energy of forward neutrons $E_{\mathrm{ZN}}$ or protons $E_{\mathrm{ZP}}$ can be measured, respectively, in neutron (ZN) and proton (ZP) Zero Degree Calorimeters (ZDCs) installed at both sides of the interaction point~\cite{Puddu2007}. The measured distributions of $E_{\mathrm{ZN}}$~\cite{Dmitrieva_ALICE_2022} demonstrated characteristic peaks corresponding to EMD events with 1,2,3...5 neutrons.  The shapes of measured ZN energy spectra are defined by the energy resolution of ZDCs~\cite{Puddu2007} and also by the efficiency of ZDCs in detecting multinucleon events~\cite{Dmitrieva2018}.
\begin{figure}[!htb]
\begin{centering}
\includegraphics[width=1.0\columnwidth]{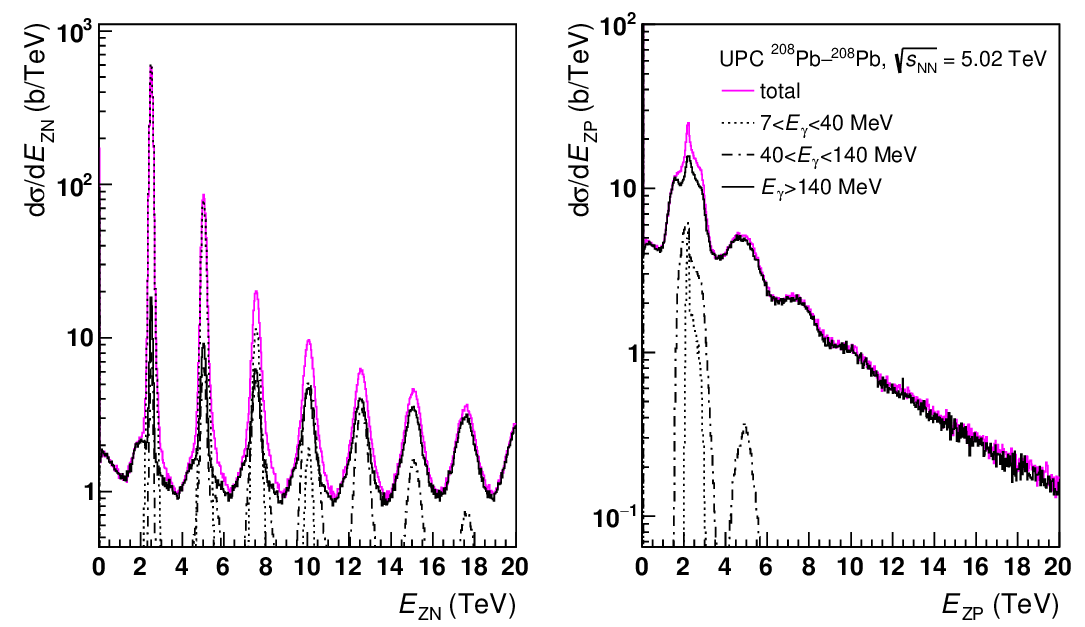}
\caption{Distributions of the total energy of neutrons (left) and protons (right) in events of EMD of $^{208}$Pb at the LHC at $\sqrt{s_{\rm NN}} = $~5.02~TeV as calculated with RELDIS. Contributions of photons of different energies are shown  as explained in the legend.  
}
\label{fig:n_p_E}
\end{centering}
\end{figure}

Since the distributions of $E_{\mathrm{ZP}}$ were not reported by ALICE, it is necessary to calculate them and compare to the $E_{\mathrm{ZN}}$ distributions to provide a guideline for future measurements of proton emission in EMD. In Fig.~\ref{fig:n_p_E} the calculated distributions of the total energy of neutrons and protons from EMD of $^{208}$Pb at $\sqrt{s_{\rm NN}} = $~5.02~TeV are shown without accounting for the energy resolution and efficiency of ZDCs. As seen, the distributions of $E_{\mathrm{ZN}}$ and $E_{\mathrm{ZP}}$ are very different because of the difference in energies of emitted neutrons and protons in the rest frame of $^{208}$Pb.  The peaks corresponding to 1n,2n,...7n emission are narrow since neutrons are mainly evaporated from an exited residual nucleus and their energies are below 1--2 MeV in its rest frame. In contrast, proton peaks are much wider, since protons are produced mainly by  photons with $E_\gamma > 140$~MeV, and thus share a noticeable part of photon energy. Due to the Lorentz boost of protons to the laboratory system, and depending on the direction of their emission, the deviation of proton energies from the beam energy of 2.51~TeV can be larger of smaller. This explains much wider proton peaks in comparison to neutron peaks and makes difficult to measure the EMD cross sections other than 1p, 2p and 3p.

\label{sec:conclusion}
\section*{Conclusion}

In this work the proton emission in EMD of $^{208}$Pb nuclei at the LHC has been studied with the RELDIS model. The contributions of different photoabsorption mechanisms to the production of elements other than Pb due to the proton emission were investigated. As shown by calculations, the cross sections of production of secondary nuclei, in particular, thallium,  can be well approximated by the cross sections of emission of the corresponding numbers of protons and neutrons. Much wider 1p, 2p and 3p peaks are expected in the distributions of the total proton energy in the EMD events in comparison to 1n, 2n, 3n peaks.


\end{document}